\documentclass[preprint,12pt]{aastex62}
\usepackage{booktabs}
\usepackage{graphicx}
\usepackage{subfigure}
\usepackage{amsmath,amssymb,latexsym,graphics,epsfig,exscale,graphicx}
\usepackage{url}
\begin{document}

\title{
ALMA Observations of Polarized 872 $\mu$m Dust Emission from the Protostellar Systems VLA 1623 and L1527
}

\correspondingauthor{Robert J. Harris}
\email{rjharris@illinois.edu}

%
%

%
%
%
\begin{abstract}
We present high-sensitivity ($\sigma_I\sim0.2-0.5$ mJy, $\sigma_{QU}\sim0.05$ mJy), high-resolution ($\sim0\farcs12-0\farcs2$) observations of polarized 872 $\mu$m dust emission from the young multiple system VLA 1623 in $\rho$ Ophiuchus and the protostar L1527 in Taurus. We detect the circumstellar material of VLA 1623A, the extended Keplerian disk surrounding VLA 1623A which we call VLA 1623CBdisk, VLA 1623B, VLA 1623W, and L1527 strongly in the polarized emission, at the $\sim1-3\%$ level. We spatially resolve VLA 1623A into two sources, VLA 1623Aa and VLA 1623Ab, separated by $\sim30$ au and located within a cavity of radius $\sim50$ au within the circumbinary Keplerian disk, as well as the edge-on disk of VLA 1623W. The polarization angle of the emission is uniform across each protostellar source and nearly  coincides with each disk's minor axis. The offsets between the minor axis and the polarization angle are not uniformly distributed at the $P\lesssim2\times10^{-4}$ level. The circumbinary disk surrounding VLA 1623Aab is azimuthally symmetrically polarized. Each compact source's emission is partially optically thick ($\tau\gtrsim1$) at 872 $\mu$m, complicating interpretations of polarization involving aligned grains. We find evidence against alignment by radiative flux in each source, particularly in the edge-on VLA 1623W and L1527. We detect astrometric offsets between the polarized emission and the total intensity in VLA 1623Aa, VLA 1623Ab, and VLA 1623B, as predicted if self-scattering in the optically thick limit operates. We conclude that self-scattering is likely responsible for disk-scale polarization at 872 $\mu$m in these systems.
\end{abstract}

\keywords{circumstellar matter--- stars: formation--- stars: magnetic field--- scattering---  individual systems: VLA 1623, L1527}
%
%

\author{Robert J. Harris}
\affiliation{National Center for Supercomputing Applications,
University of Illinois at Urbana-Champaign, 1205 W Clark St, Urbana, IL 61801} 

\affil{Department of Astronomy, University of Illinois at Urbana-Champaign,
1002 W. Green St.,
Urbana, IL 61801, USA}

\author{Erin G. Cox}
\affil{Department of Astronomy, University of Illinois at Urbana-Champaign,
1002 W. Green St.,
Urbana, IL 61801, USA}

\author{Leslie W. Looney}
\affil{Department of Astronomy, University of Illinois at Urbana-Champaign,
1002 W. Green St.,
Urbana, IL 61801, USA}

\author{Zhi-Yun Li}
\affiliation{Department of Astronomy, University of Virginia, 530 McCormick Rd, Charlottesville, VA 22903, USA}

\author{Haifeng Yang}
\affiliation{Department of Astronomy, University of Virginia, 530 McCormick Rd, Charlottesville, VA 22903, USA}

\author{Manuel Fern\'andez-L\'opez}

\affiliation{Instituto Argentino de RadioastronomÂŽÄ±a (CONICET), CCT La
Plata, 1894, Villa Elisa, Argentina}

\author{Woojin Kwon}
\affiliation{Korea Astronomy and Space Science Institute, 776 Daedeok-daero, Yuseong-gu, Daejeon 34055, Korea}
\affiliation{Korea University of Science and Technology, 217 Gajang-ro, Yuseong-gu, Daejeon 34113, Korea}

\author{Sarah Sadavoy}
\affiliation{Harvard-Smithsonian Center for Astrophysics, 60 Garden St, Cambridge, MA 02138, USA}

\author{Dominique Segura-Cox}
\affiliation{Max Planck Institute for extraterrestrial Physics, Giessenbachstrasse 1,
85748 Garching,
Bayern,
Deutschland
}

\affil{Department of Astronomy, University of Illinois at Urbana-Champaign,
1002 W. Green St.,
Urbana, IL 61801, USA}

\author{Ian Stephens}
\affiliation{Harvard-Smithsonian Center for Astrophysics, 60 Garden St, Cambridge, MA 02138, USA}

\author{John Tobin}
\affiliation{Homer L. Dodge Department of Physics and Astronomy,  Oklahoma University, 440 W. Brooks St.
Norman, OK 73019, USA}
\affiliation{Leiden Observatory, Leiden University, P.O. Box 9513, 2000-RA Leiden, The Netherlands}

%
%
%

\section{Introduction}  \label{sec:intro} 

Magnetic fields play a key role in the formation and evolution of protostellar objects and their associated disks. In the early stages of protostellar formation, on $\sim$ pc-scales, magnetic fields help regulate the collapse via ambipolar diffusion \citep[e.g.,][]{sh87} and can funnel material from large-scales to small-scales \citep{li14a}. On small ($\sim$ few au) scales, magnetic fields are crucial elements of outflows, jets, wind launching, and disk accretion \citep[e.g.,][]{bl82,ba98}. Theoretical studies have suggested that magnetic fields can efficiently brake infalling material and limit disk sizes to $r\lesssim 10$ au during the Class 0 \citep{an93} phase.
Simulations in ideal MHD \citep[e.g.,][]{me08,he08,jo12} show that this braking is most efficient if the field and the angular momentum of the material are aligned, whereas the braking becomes less efficient as the orientation changes from parallel to perpendicular \citep{me08,he08,li11,jo12}. If the material efficiently loses angular momentum, large ($r \gtrsim 100$ au) disks can only develop during the Class I and II protostellar stages (e.g., \citealt{me09,an09,an10,da10}); if it is not, large disks can form earlier. Simulations that include non-ideal MHD effects \citep[e.g., ][]{ma16,he16,va18} indicate that disks with $r >$ 50 au might be able to form even if the magnetic field is aligned the rotation axis, although the bulk of the duration of the Class 0 phase was not simulated in these studies.


Some observational support for the significance of magnetic braking has come from large-scale ($\sim 1000$ au) observations of relative orientations of outflow axes and magnetic field orientations inferred from millimeter-wave dust polarization in young Class 0 objects \citep[e.g.,][]{hu14,se15}. These studies typically find that disk-bearing sources have orthogonal orientations, whereas sources with no evidence of disks larger than $\sim 10$ au have roughly parallel orientations. It is, however, unclear that the dust polarization provides a good tracer of the magnetic field orientation as one gets closer into the protostar ($r \lesssim 100$ au) where physical conditions are vastly different than those in the outflow or outer envelope.

Constraining how magnetic fields guide early disk evolution is predicated on finding a good observational proxy for the magnetic field that can be mapped at high spatial resolution near the protostar. It has traditionally been assumed that, at sub/millimeter-wavelengths, the origin of polarized continuum is dust grains aligned with the local magnetic field \citep{la07}. In this scenario, non-spherical dust grains align their short axes with the local magnetic field, creating an axis-dependent optical depth. In the optically thin limit, this alignment yields thermal emission polarized perpendicular to the local magnetic field. Such an origin would be useful to study the magnetic field structure that accompanies protoplanetary disk formation.



Another plausible origin for the polarized continuum emission is self-scattering of the thermal dust emission. Light of wavelength $\lambda$ is effectively scattered by dust grains of size $a \sim \lambda/(2\pi)$ and yields polarized emission in the presence of a radiation anisotropy. The resulting emission is polarized orthogonal to the direction of anisotropy in the local radiation field at the same wavelength. Studies have recently shown that self-scattering can be quite effective, even at sub/millimeter wavelengths or longer, at generating $\sim$ few percent levels of polarized emission in protostellar/planetary disks with predictive morphologies that depend on disk inclination \citep{ka15,ka16,ya16a,ya16b}. Due to the steep dependence of the scattering cross-section on the ratio $a/\lambda$, the polarized scattered emission has been suggested as a good way to directly probe the grain-size distribution independent of the opacity spectral index $\beta$, determination of which is complicated by finite optical depth as well as the probability of being out of the Rayleigh-Jeans limit at shorter wavelengths \citep{ka15,ka16}.

A third mechanism that may contribute to the polarization signal is that from thermal emission from grains aligned with the momentum gradient of the local radiation field. This scenario differs from the magnetically aligned scenario in that the (predominantly mid/far-infrared) stellar light is re-processed through the disk and strikes the ellipsoidal grains. These reach an equilibrium when their cross-section of largest area is perpendicular to the local radiation field anisotropy \citep{ta17}. The 3 mm continuum polarization structure of the disk of the Class I/II source HL Tau has been attributed to this mechanism \citep{ka17}.

Each of these scenarios predicts polarization angle and polarization intensity distributions in young protostellar disks where grain sizes are large enough for both scattering and grain alignment to be efficient. For optically thin emission from magnetically aligned grains, the polarization is orthogonal to the local magnetic field as projected onto the plane of the sky. So a toroidal magnetic field in a disk viewed face-on would have a radial polarization pattern, whereas a poloidal field with a radial component in the same source would have a roughly azimuthal pattern. Inclination can change these morphologies, particularly at the extremes: an edge-on disk threaded by a toroidal field would have polarization parallel to the disk's minor axis. In the case of scattering, the polarization morphology and strength depend on the details of the morphology of the radiation field and the disk inclination angle. Typically, the polarization is along the projected minor axis of the inclined disk, especially in the central region, although it can become more azimuthal in the optically thin outer part of the disk where the local radiation field is beamed radially outward \citep[e.g.][]{ya16a,ka16}. In the case of grain alignment by the momentum gradient of the radiation field, the polarization direction at each location in the plane of the sky would be perpendicular to the local radial direction (\citealt{ta17}, Yang et al., in prep). Inclination, optical depth, and beam-depolarization effects can alter the degree to which these morphologies are recoverable. 

Recent observational work to disentagle these mechanisms in young systems has yielded considerable insights. For instance, high-resolution, multi-band Atacama Large Millimeter/submillimeter Array (ALMA) observations of the Class I/II system HL Tau by \citet{ka17} and \citet{st17b} have indicated that the polarization pattern may evolve rapidly with changing wavelength. At 870 $\mu$m, HL Tau's polarization is predominantly along the minor axis, strongly suggesting that scattering dominates the polarized emission at this band, while the polarization pattern is azimuthal at 3 mm, suggesting that alignment by radiative flux is responsible at this band. \cite{co18} observed 10 Class 0 systems in Perseus as part of follow-up to the VLA Nascent Disk and Multiplicity (VANDAM) survey \citep{to15,to16} to image the 870 $\mu$m dust continuum polarization. Three of these sources are candidate disk systems -- i.e., their 8 mm visibility profile is well fit by a viscously-evolving self-similar disk profile \citep{se16,se17} -- while seven are not. \citeauthor{co18} found that the disk candidates preferentially have small-scale (within about 100 au) polarization preferentially aligned with the minor axis of the disk (and the outflow axis), while non-disk candidates have small-scale polarization randomly oriented with respect to the outflow axis. These results suggest that scattering may be responsible for the small-scale polarization at sub/millimeter-wavelengths in the disk systems. On larger scales, the more extended sources for which significant envelope was detected showed complicated morphologies that suggest a significant contribution from magnetically aligned grains in the envelope. \citet{le18} find a similar alignment between the minor axis and the 870 $\mu$m continuum polarization angle in the disk of the young Class 0 system HH 212 and in the inner disk of the Class 1 system HH 111 VLA 1, with a transition to a more radial orientation in the outer disk. \citet{gi18} find a complicated morphology to the 1.14 mm continuum polarization of HH 80-81, with a sharp transition in morphology similar to what is observed in HH 111 VLA1. They attribute the abrupt change in morphology to the transition between optically thick and optically thin emission.

To further understand the origin of the polarized disk emission, we have obtained high-resolution ($\sim 0\farcs1$), high-sensitivity ($\sigma \lesssim 0.2-0.5$ mJy) polarimetric observations of two young Class 0 systems, VLA 1623 and L1527 from ALMA in the 872 $\mu$m continuum. These sources represent ideal targets for studying disk polarization because they are relatively nearby (VLA 1623, $d\sim 137$ pc, \citealt{or17}; L1527, $d\sim 145$ pc, \citealt{lo07}), are known to harbor kinematically-confirmed Keplerian disks \citep{mu13,to12,oh14}, and are bright in the sub/millimeter bands ($F_{850\mu m} \gtrsim$ few 100s of mJy). 

The VLA 1623AB+W system is in Ophiuchus and comprises the inner VLA 1623AB Class 0 binary with separation $\sim 1\farcs2$ along with a more spatially and kinematically separated Class I component, VLA 1623W, at a distance $\sim 10\farcs5$ and kinematic offset $3-4$ km/s from the VLA 1623AB subsystem \citep{bo97,lo00,ma12,mu13}. The Keplerian disk in this system orbits the easternmost A component and suggests a component mass of 0.1-0.2 M\textsubscript{\(\odot\)} \citep{mu13}. All three objects have been strongly detected in Band 6 (1.3 mm) continuum with ALMA \citep{mu13}. Large-scale, $\sim$ 3 arcsecond ($\sim 400$ au) resolution 1.3 mm continuum polarimetry with the Combined Array for Millimeter Astronomy (CARMA) has shown that the inferred magnetic field orientation is misaligned with the outflow axis, consistent with a lack of magnetic braking that the large Keplerian disk suggests \citep{mu13,hu13,hu14}. Recent $\sim 0\farcs3$ resolution observations of the Band 6 continuum by \cite{sa18} show strong linear polarization of both the A and B components with position angles roughly aligned with the outflow axis of the system. This data also show significant azimuthally polarized emission towards the extended circumstellar disk surrounding VLA 1623A -- VLA 1623CBdisk (for circumbinary disk; see Section \ref{sec:results}). They attribute the bulk of the polarization in the extended disk to magnetically aligned grains and that in the compact protostellar sources VLA 1623A and VLA 1623B to self-scattering. 

The L1527 system is an isolated Class 0 object in Taurus. The system's Class 0 disk has been resolved in the 0.87 and 3 mm continuum with CARMA \citep{to13} and has been kinematically confirmed to be Keplerian, orbiting a central mass of between $\sim$ 0.2 and 0.45 M\textsubscript{\(\odot\)} \citep{to12,oh14,as17}. CARMA observations at 3 arcsecond resolution showed the 1.3 mm polarization to be roughly parallel to the outflow axis \citep{hu14}, again consistent with the overall picture of an inferred magnetic field perpendicular to the outflow axis and the presence of a large, kinematically confirmed disk \citep{to12}. Higher resolution ($0\farcs3$) CARMA observations at 1.3 mm showed that this configuration holds even at $\sim$ 50 au scales and that the disk is polarized at the $\sim 2-3\%$ level with a polarization structure consistent with an origin in grains aligned with a toroidal magnetic field \citep{se15}. \cite{da14} have used multi-band, multi-scale polarization observations and theoretical modeling to show that the extant observations and presence of a large disk are consistent only with a collapse that started with a weak magnetic field oriented orthogonal to the outflow axis.

In this paper, we present our ALMA observations of these two systems. In Section 2, we present our observational setup for the two sets of observations; in Section 3; we present our images of hte two systems and summarize the measurements of the sizes and flux densities (both in Stokes I and in polarized intensity) for the systems; and in Section 4, we discuss our observations in the context of mechanisms for producing polarized emission in young sources; we also discuss how our observations inform how the system VLA 1623 formed. 

\section{Observations}  \label{sec:obs}

Our ALMA Cycle 3 observations (project code: 2015.1.00084.S) were conducted on 21 August 2016 (for VLA 1623) and 12 September 2016 (for L1527). Observations
of VLA 1623 were conducted in the ALMA configuration C40-5 with baselines ranging from 14 meters to 1.45 km, yielding a typical spatial resolution of $\sim 0\farcs14$ and an estimated maximum recoverable scale of $\sim 2\farcs3$, while those of L1527 were conducted in the
ALMA configuration C40-6 with baselines ranging from 21 m to 3.14 km, yielding a typical spatial resolution of $\sim 0\farcs09$ and an estimated maximum recoverable scale of $\sim 1\farcs4$ at 345 GHz. Both our VLA 1623 observations and our L1527 observations comprised a single pointing each. During the VLA 1623 observations, we observed both the Class 0 multiple VLA 1623AB and its $\sim$ 10$''$ companion, the Class I source VLA 1623W, by selecting a pointing center equidistant to both sources. Each source was $\sim 5''$ from the pointing center. This places both sources outside the inner $\sim$ 1/3 of the primary beam full-width at half-max, $\sim 3''$ radius at 872 $\mu$m, over which the ALMA Science Center indicates that polarimetry is reliable. However, we are confident in the fidelity of our polarimetric observations for three reasons. First, the polarization morphology of VLA1623AB matches well that found at Band 6 by \cite{sa18} which were centered on VLA 1623 AB. Furthermore, the polarization morphology of VLA 1623W is also consistent with Band 6 observations centered on this source, (Sadavoy et al. in preparation). Finally, ALMA commissioning tests at Band 7 show that reliable polarimetry (systematic polarization fraction accuracy $\lesssim 0.5$\% and systematic polarization angle accuracy $\lesssim 1$ degree) can be attained for strongly polarized sources such as ours at the distance our sources are away from the pointing center by doing the standard `on-axis' polarimetric calibration. (P. Cortes, private communication).

The ALMA observations were conducted at Band 7, with the local oscillator set to $\nu_{LO} = 343.5$ GHz,
with the two spectral windows in the lower sideband centered at 336.4945750 and 338.4320750 and the two spectral windows in the upper sideband centered at 348.4945750 and 350.4945750 GHz. Each spectral window was divided into 64 channels of 31.25 MHz apiece, yielding a total bandwidth per spectral window of 2 GHz. For each
channel/baseline pair, all four possible correlations, XX, YY, XY, and YX, were formed to allow for polarimetric imaging. 

The observations of the VLA 1623 system included the quasars J1517-2422 as the flux and bandpass calibrators, J1625-2527 as the complex gain calibrator, and J1633-2557 as the
check source to establish the fidelity of calibration transfer from the gain calibrator to the source. J1512-0905 was used to calibrate the cross-hand delay, X-Y phase offset, and the D-terms (or leakage). 
The observations of the L1527 system included J0510+1800 as the bandpass
and flux calibrator, J0433+2905 as the complex gain calibrator, and J0440+2728 as the check source. J0510+1800 was used to calibrate the cross-hand delay, X-Y phase offset, and the D-terms. Observations of the target source were interleaved with observations of both the complex gain calibrator as well as the check source. 

These observations were reduced manually by analysts at the National Radio Astronomy Observatory (NRAO), using
the Common Astronomy Software Applications (CASA) package \citep{mc07}, version 4.7.0. Briefly, a priori calibrations were first applied to the data. These included the system temperature calibration (to establish a pseudo-flux scaling as well as to calibrate the data weights) and an a priori phase calibration from water vapor radiometer measurements. After these calibrations, the data were bandpass calibrated, gain calibrated, and then flux calibrated using the usual bootstrap via the initial gain solutions. The amplitude scales for the observations were set by the flux calibrators. Quasar amplitude calibrators were used for the calibration. The flux densities for these calibrators were established using the ALMA Calibrator Database. The assumed flux density for J0510+1800 at 343.425 GHz was 2.238 Jy with a spectral index of -0.2, while that for J1517-2422 was 2.885 Jy with a spectral index of -0.15. The overall amplitude calibration uncertainty at Band 7 for ALMA is estimated at 10\%.

After the calibration of the parallel-hand correlations, the cross-hand correlations were calibrated. First, the polarization, flux, and bandpass calibrator model was reset. The data were gain calibrated again assuming zero source polarization, and the variation in the inferred X and Y gain amplitudes for each antenna used to establish an estimate of the source polarization. Second, the calibrator scan for which the cross-hand signal is maximized (and parallel-hand signal minimized) was used to calibrate the cross-hand delay (i.e., the phase variation along the bandpass for the cross-hand signal). Third, the absolute phase offset between the X and Y polarizations as well as the source polarization properties were determined. Because there is a phase ambiguity of 180 degrees in the source polarization when solving for both the phase offset and the source polarization, estimates determined from the initial gain calibration mentioned above were used to resolve the ambiguity. Finally, the D-terms were determined. 

After both parallel- and cross-hand correlations were calibrated, we averaged the data in frequency to form one 2 GHz wide channel per spectral window. The science targets were then \texttt{split} out and imaged. The data were imaged in all four Stokes parameters in CASA 4.7.0 using the \texttt{clean} task with \texttt{beammode=`clarkStokes'} to ensure optimal deconvolution in all four Stokes parameters as well as using the \texttt{multiscale} parameter to ensure that both large- and small-scale emission were properly deconvolved. Both the VLA 1623 and L1527 systems were sufficiently bright to self-calibrate, so we performed a phase-only self-calibration on them, solving for the antenna-based phases every $\sim 6$ seconds. For this work, we imaged the L1527 data using natural weighting, which resulted in a beam size of $0\farcs23 \times 0\farcs14$ (PA = -3.9$^\circ$). For the VLA 1623 system, we produced both a naturally weighted image with a beam size of $0\farcs 20 \times 0\farcs17$ (PA = 57.1$^\circ$) and a Briggs-weighted image with robust = -2 (i.e., close to uniform weighting), for which the beam was $0\farcs14 \times 0\farcs12$ (PA = 29.8$^\circ$). Because of the locations of each of the components in the VLA 1623 system within the primary beam, these images were corrected for the ALMA primary beam pattern as well, yielding a $\sim$ 20\% correction in the surface brightness of each component. Due to its proximity to the pointing center, no primary beam correction was applied to the L1527 images. A summary of the images produced is given in Table 1. 

\begin{deluxetable}{cccccccccc}
\label{tab:image}
\tablecaption{Summary of imaging}
\tablenum{1}
\tablehead{\colhead{Star} & \colhead{Weighting} & \multicolumn{3}{c}{Synthesized beam} & \multicolumn{5}{c}{$\sigma_{I,Q,U,V,PI}$} \\ 
\colhead{} & \colhead{} & \multicolumn{2}{c}{(arcsec)}  & \colhead{(degrees)} & \multicolumn{5}{c}{(mJy/beam)}  } 
\startdata
VLA 1623AB &  natural &   0.20 &  0.17 &  57.1 &  0.55 &  0.044 &   0.079 &  0.059 &  0.042 \\
VLA 1623AB &  Briggs &   0.14 &  0.12 &  29.8 &  0.19 &   0.053 &   0.057 &   0.054 &  0.036 \\
VLA 1623W &  natural &   0.20 &  0.17 &  57.1 &  0.32 &   0.043 &  0.055 &   0.127  &  0.050 \\
L1527         &  natural &   0.23 &  0.14 &  -3.9 &   0.17 &  0.036 &  0.031 &  0.032 &  0.023 \\
\enddata
\end{deluxetable}

Our calibrated Stokes cubes were used to make polarization intensity (PI), polarization angle (PA), and polarization fraction maps. We used
the CASA task \texttt{immath} to do this. The linear polarization intensity map represents the quantity $PI = \sqrt{Q^2+U^2}$, and the polarization angle map represents the quantity PA $= 0.5 \arctan(U/Q)$. The PI maps were debiased using the average noise value determined from the Q and U maps. Each of the PI, PA, and polarization fraction maps was masked below 5$\sigma$ in PI. The linear polarization fraction map was then formed by dividing the PI map by the total Stokes I map. The Stokes I map was masked below $3\sigma$ for this division, so each polarization map shows where there is emission above both 3$\sigma$ in Stokes I and 5$\sigma$ in polarized intensity.

\section{Results}  \label{sec:results}

\begin{figure*}[htp]
\centering
\includegraphics[width=\textwidth]{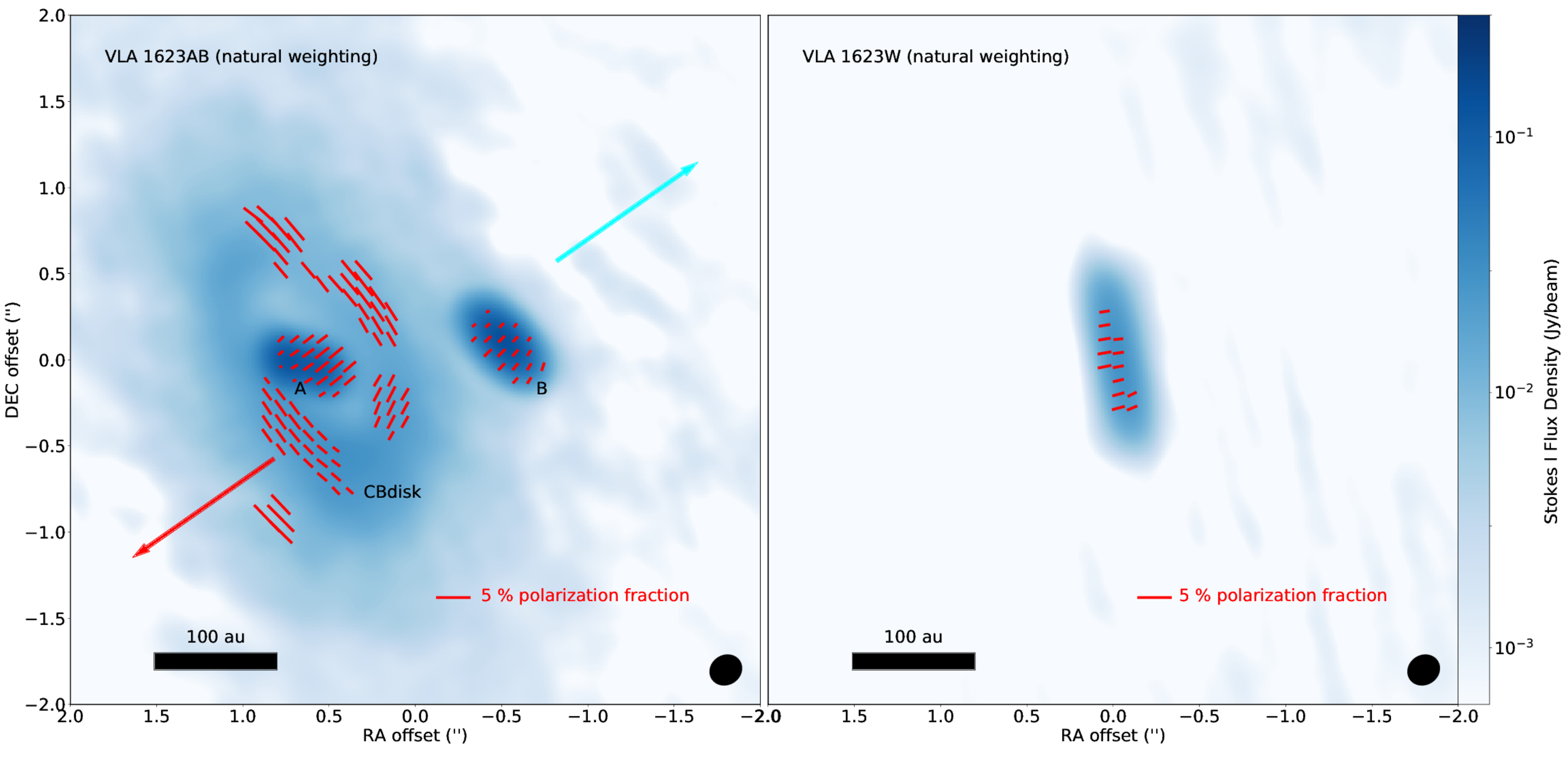}
\caption{(\textit{left}) Naturally weighted image of the 872 $\mu$m continuum emission from the Class 0 triple system VLA 1623AB. The A and B protostellar components are labeled, as is the circumbinary reservoir. The outflow orientation is indicated by red and blue arrows. (\textit{right}) same image for the $\sim 10''$ Class 0 companion VLA 1623W. The scale-bars indicate 100 au. The synthesized beam is indicated in the lower right. Horizontal bars in the middle of the panels indicate the length of a 5\% polarization vector. The colorbars give the surface brightness for both images.}
\label{fig:vla1623}
\end{figure*}

\begin{figure*}[htp]
\centering
\includegraphics[scale=0.28]{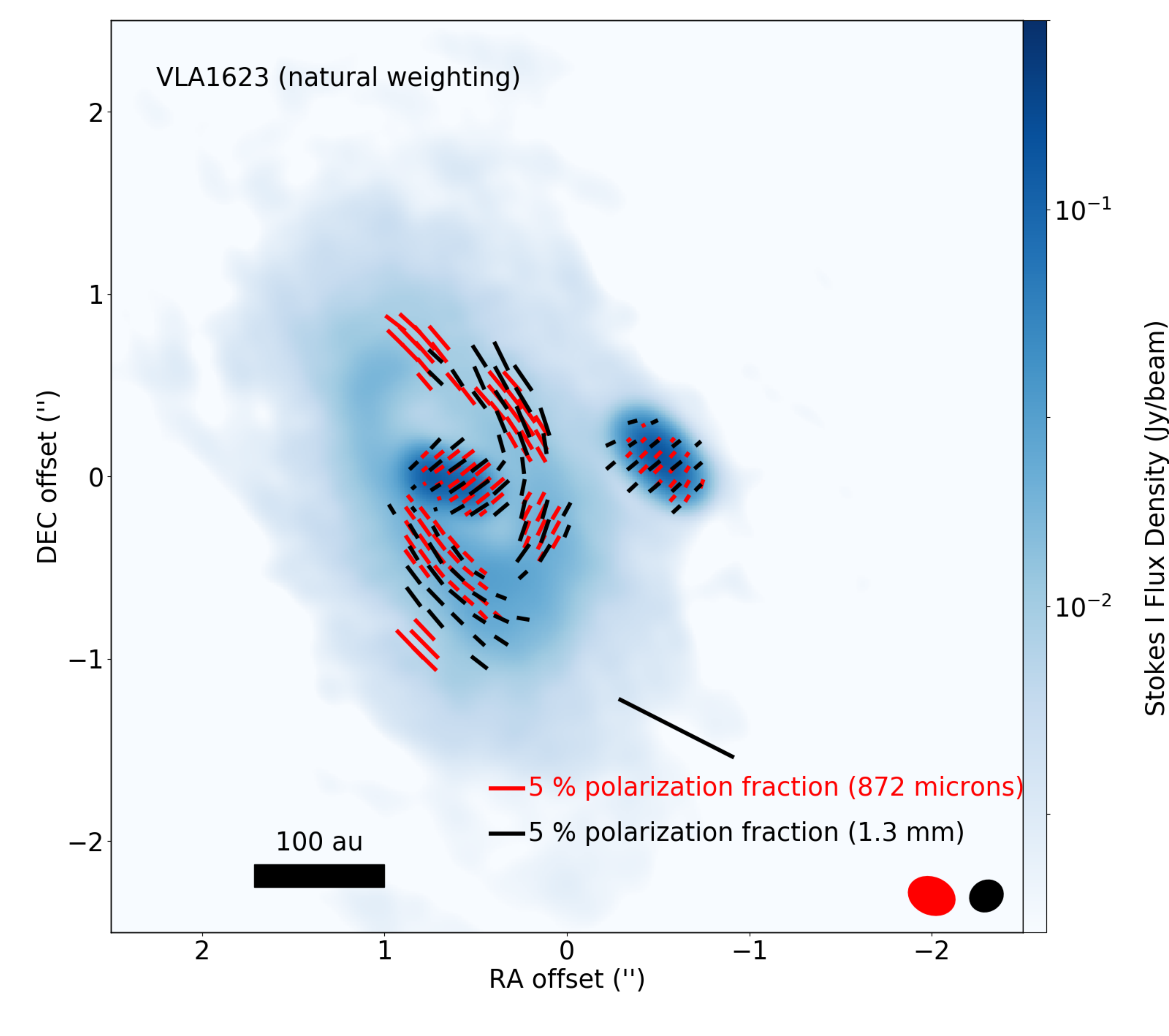}
\caption{Comparison of the polarization morphologies of the dust continuum emission in VLA 1623AB from our 872 $\mu$m observations (Stokes I emission in colorscale, polarization angle/fraction in black) and from the 1.3 mm observations (polarization angle/fraction in red) by \cite{sa18}. The synthesized beams are indicated in the lower right.}
\label{fig:vla1623comparison}
\end{figure*}

\begin{figure*}[htp]
    \centering
    \subfigure[]{\includegraphics[scale=0.20]{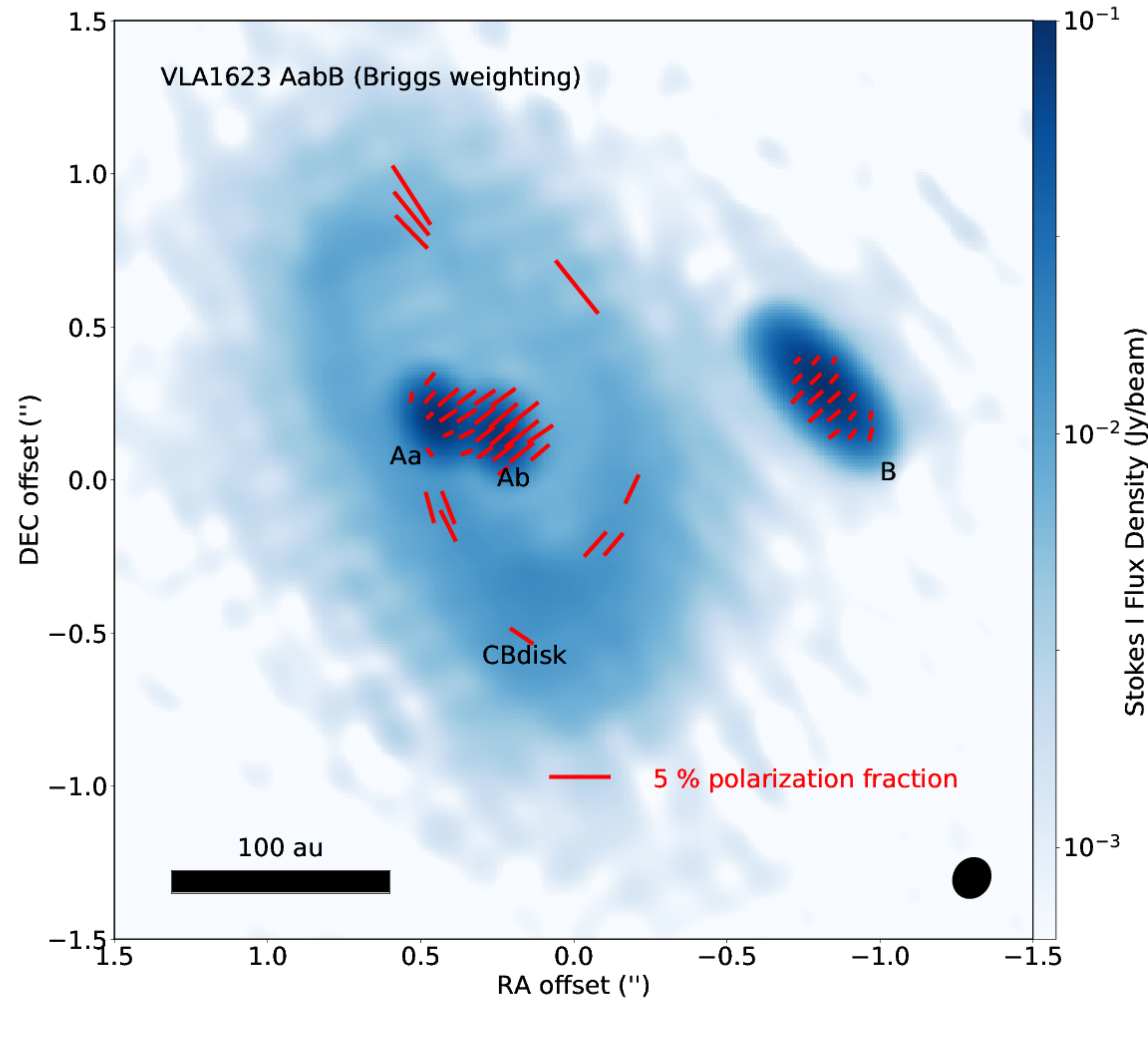}}
    \subfigure[]{\includegraphics[scale=0.20]{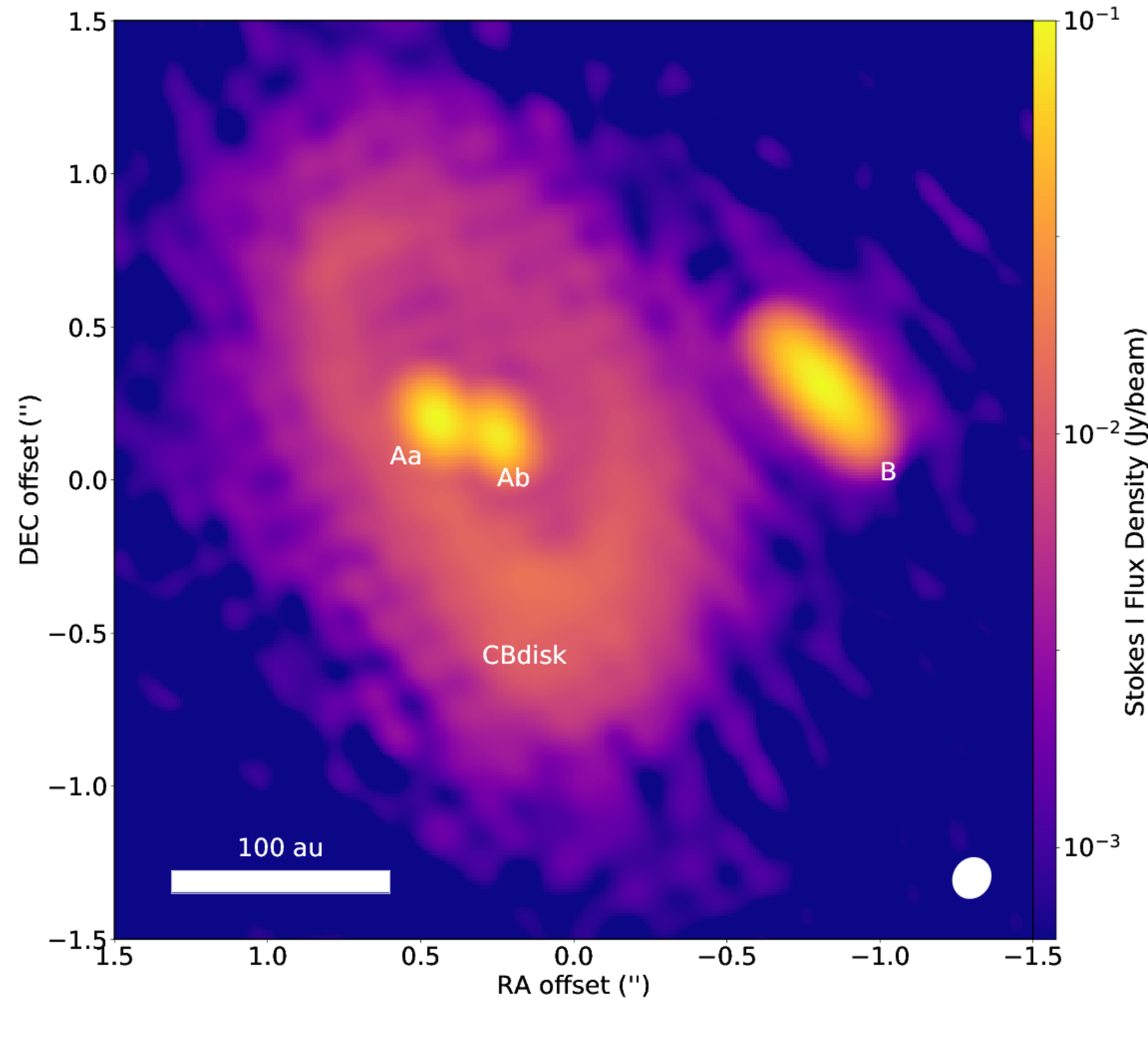}}
\caption{\textit{(left)} Briggs-weighted image of the 872 $\mu$m continuum emission from the Class 0 system VLA 1623AabB. The scale-bar indicates 100 au. The synthesized beam is indicated in the lower right. The horizontal bar in the middle indicates the length of a 5\% polarization vector. The colorbar to the right gives the surface brightness. \textit{(right)}: Identical image to left, but with a different color mapping to better show the compact sources VLA 1623Aa and VLA 1623Ab.}
\label{fig:vla1623r}
\end{figure*}

\begin{figure*}[htp]
\centering
\includegraphics[scale=0.28]{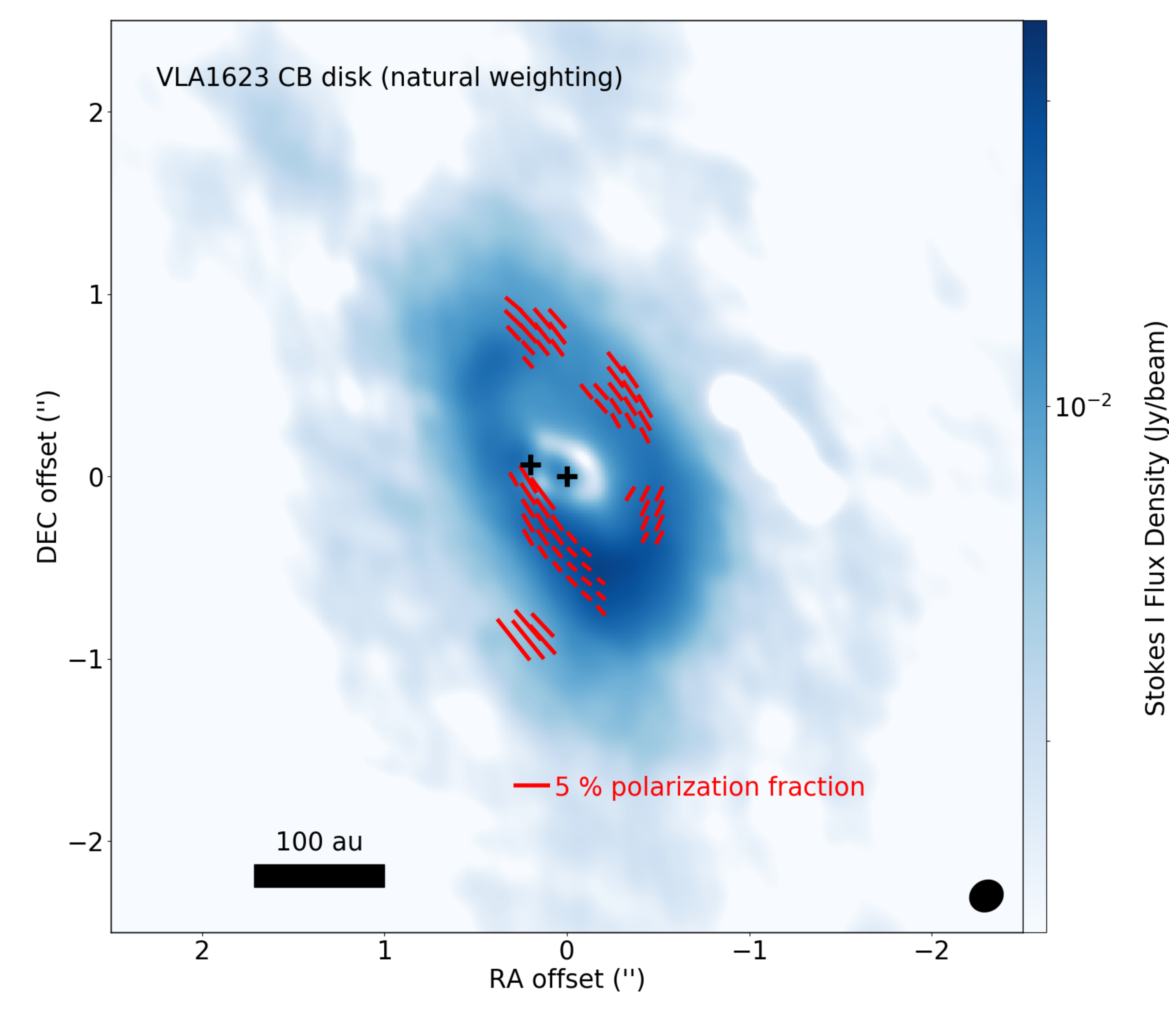}
\caption{ Naturally weighted image of the 872 $\mu$m continuum emission from the circumbinary disk surrounding VLA 1623Aab. The best fit components to the Briggs-weighted map from each of Aa, Ab, and B were subtracted from the visibility data prior to imaging. The stellar positions of Aa and Ab are marked with crosses. The scale-bar indicates 100 au. The synthesized beam is indicated in the lower right. The horizontal bar in the middle indicates the length of a 5\% polarization vector. The colorbar to the right gives the surface brightness}
\label{fig:vla1623cb}
\end{figure*}

\begin{figure*}[htp]
\centering
\includegraphics[scale=0.28]{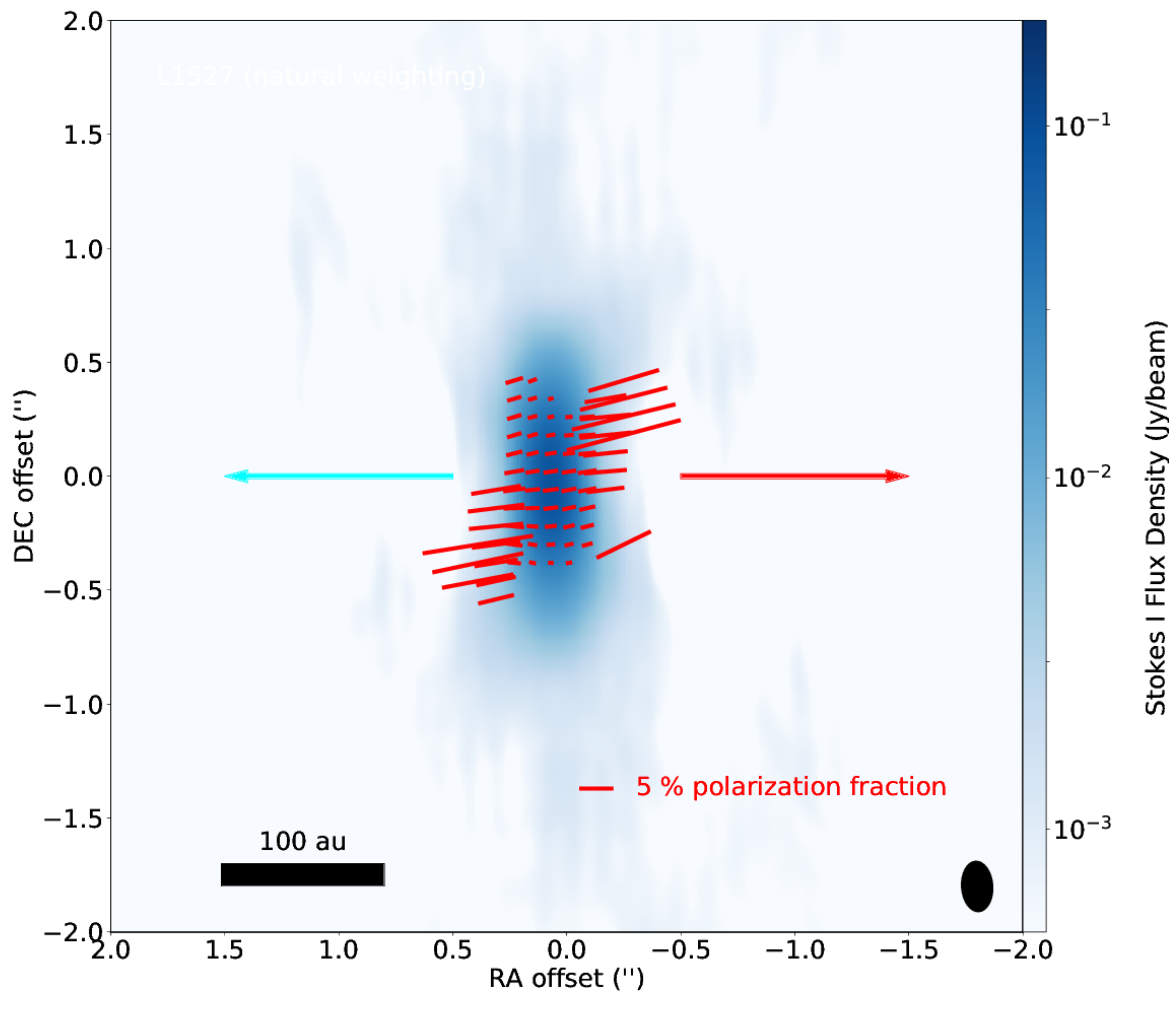}
\caption{Naturally weighted image of the 872 $\mu$m continuum emission from the Class 0 system L1527. The outflow orientation is indicated by red and green arrows. The scale-bar indicates 100 au. The synthesized beam is indicated in the lower right. The horizontal bar in the middle indicates the length of a 5\% polarization vector. The colorbar to the right gives the surface brightness}
\label{fig:l1527}
\end{figure*}

We show the images of the VLA 1623 multiple system in Figures \ref{fig:vla1623}, \ref{fig:vla1623comparison}, \ref{fig:vla1623r}, and \ref{fig:vla1623cb}, and that of the L1527 system in Figure \ref{fig:l1527}. The properties of each system as measured from the different images are shown in Table 2.
These properties are determined from image-plane fitting of elliptical Gaussians to each of the objects in the VLA 1623 and L1527 maps, both in total intensity and in the polarized intensity. All fits presented here are deconvolved from the synthesized beam.

\begin{deluxetable}{rrrrrrcccc}
\label{tab:properties}
\tabletypesize{\tiny}
\setlength{\tabcolsep}{1pt}
\tablecaption{Summary of image plane fitting}
\tablenum{2}
\tablehead{\colhead{Weighting} & \colhead{Object} & \colhead{Right ascension} & \colhead{Declination} & \colhead{Ellipse size} & \colhead{Ellipse PA} & \colhead{Peak I} & \colhead{Int. I} & \colhead{Peak PI} & \colhead{Int. PI} \\ 
\colhead{} & \colhead{} & \colhead{(J2000)} & \colhead{(J2000)} & \colhead{(milliarcsec)} & \colhead{(degrees)} & \colhead{(mJy/beam)} & \colhead{(mJy)} & \colhead{(mJy/beam)} & \colhead{(mJy)}} 
\startdata
Natural & VLA 1623Aab &  16.26.26.3933 $\pm$ 0.0008  &  -24.24.30.828 $\pm$ 0.005 &  393   ($\pm$ 31) $\times$ 157 ($\pm$ 18)  &  76 $\pm$ 3    &   121 $\pm$ 7      & 368 $\pm$ 26   & 2.33 $\pm$ 0.12 &  4.79 $\pm$ 0.36	 \\
 &VLA 1623B   &  16.26.26.3066 $\pm$ 0.0001   &  -24.24.30.694 $\pm$ 0.002 &  335  ( $\pm$  6 )$\times$ 107 ($\pm$ 4)  &  41.6 $\pm$ 0.6 &   139 $\pm$ 2      &  324$\pm$ 5    & 1.94 $\pm$ 0.02 &  2.95 $\pm$ 0.04	 \\
& VLA 1623W   &  16.26.25.6313 $\pm$ 0.0002  &  -24.24.29.589 $\pm$ 0.010   &  828.1 ($\pm$ 26) $\times$ 120 ($\pm$ 8)  &   10 $\pm$ 0.5  &   29 $\pm$ 1 &  159$\pm$ 5  & 0.427$\pm$ 0.03 &  2.65 $\pm$ 0.2	 \\
 & L1527      &  04:29:53.8753 $\pm$	0.0001  &  +26.03.09.511 $\pm$ 0.008 &  581.1 ($\pm$	19) $\times$ 148	($\pm$ 5)  & 	1.8 $\pm$	0.6  &   122  $\pm$ 3  &  488 $\pm$ 14 &  2.65	$\pm$ 0.12 &  7.80  $\pm$ 0.50 \\
Briggs &VLA 1623Aa  &  16.26.26.3981 $\pm$ 0.0002  &  -24.24.30.807 $\pm$ 0.003 &  158 ($\pm$ 13)   $\times$  108 ($\pm$ 11) &  48 $\pm$ 11   &  96 $\pm$ 4        &  193$\pm$ 10   &  1.46 $\pm$ 0.30  &  1.46 $\pm$ 0.30	 \\
 & VLA 1623Ab  &  16.26.26.3847 $\pm$ 0.0002  &  -24.24.30.854 $\pm$ 0.002 &  150 ($\pm$ 9)    $\times$  93  ($\pm$ 18) &  49 $\pm$ 6    &  83 $\pm$ 2        &  152$\pm$ 6    &  2.21 $\pm$ 0.03 &  2.93 $\pm$ 0.05 \\
 & VLA 1623B   &  16.26.26.3065 $\pm$ 0.0001  &  -24.24.30.693 $\pm$ 0.002 &  336 ($\pm$ 6)    $\times$  94  ($\pm$ 3)  &  41.4$\pm$ 0.5 &  98 $\pm$ 1        &  321$\pm$ 5    &  1.40 $\pm$ 0.03 &   2.3 $\pm$ 0.10 \\
100k$\lambda$ & VLA 1623CBdisk$^{\dagger}$ & $ $ & $ $ & 2166 ($\pm$ 99) $\times$ 1221 ($\pm$ 83) & 25.2 $\pm$ 3.8 & 378 $\pm$ 12 & 1096 $\pm$ 46 &  \\
\enddata
\tablecaption{Results of fitting sources to elliptical Gaussians. $\dagger$}
\tablecomments{All uncertainties are statistical. The estimated systematic amplitude uncertainty of 10\% is not reflected in the quoted uncertainties. Because of beam-depolarization, \texttt{imfit} could not be used to estimate the total polarized intensity for VLA 1623CBdisk.}
\end{deluxetable}

The naturally weighted image of VLA 1623AB (Figure \ref{fig:vla1623}, left) shows the continuum emission from both the A and B components, as well as that from the circumstellar material that surrounds VLA 1623A. Each component of the system is at least marginally resolved and also strongly detected in polarized continuum. The individual protostellar components are polarized at the $\lesssim$ 1 \% level in integrated emission; polarization of the surface brightness reaches $\sim$ 1-3\% levels. The polarization strength and morphology of the A and B sources are consistent with those found by \cite{sa18} at Band 6 (1.3 mm; see Figure \ref{fig:vla1623comparison}). Likewise, the circumstellar reservoir that surrounds A is also detected and resolved in the polarized continuum, as first reported in the observations by \citeauthor{sa18} We find the emission from the ring to be azimuthally polarized at the $\sim 2-5\%$ level, in agreement with the 1.3 mm continuum results. While the polarization angle is consistent throughout most of the ring between the two images, the patch of polarized emission we recover at a position angle $\sim$ -40 degrees (north towards east) has a slightly more azimuthal orientation than the corresponding patch at 1.3 mm.
 The angular difference is $\sim 10^{\circ}$. This is the only region of VLA 1623 where the polarization morphology recovered from the data significantly differs between 872 $\mu$m and 1.3 mm (see Section \ref{ref:polsubsection}). When the data are combined, we find that the spectral index of both VLA 1623A and B (excluding the extended Keplerian disk orbiting VLA 1623A) is approximately $\alpha \sim 2.30 \pm 0.45$ assuming a 10\% absolute calibration uncertainty at each of the two bands. This is consistent with the determination of \citeauthor{sa18} that the emission from these sources is optically thick. 

The naturally weighted image of VLA 1623W (Figure \ref{fig:vla1623}, right) shows for the first time a clearly resolved, extended disk-like object in this Class I source. Similarly to the other protostellar components, the circumstellar emission is polarized at the $\lesssim$ 1 \% level in integrated emission, with polarization of the surface brightness reaching $\sim$ few \% levels.

The higher resoluton ($\sim 0\farcs12$) Briggs-weighted map of VLA 1623AB (Fig. \ref{fig:vla1623r}) shows clearly that the A component separates into two individual components, the brighter VLA 1623Aa in the east and dimmer VLA 1623Ab in the west. This is the first time that the compact emission associated with VLA 1623A has been resolved. Given the resolution of these two components as well as the gap-like structure in the circumstellar reservoir surrounding them, it is likely that they represent a proto-binary ($\rho = 0\farcs21 \sim 30$ au) within the VLA 1623AB system. If true, this would  make VLA 1623AB a likely protostellar hierarchical triple system. Interestingly, while VLA 1623Aa is brighter in total emission than VLA 1623Ab (193 vs 152 mJy), it is dimmer in the polarized intensity (1.46 vs 2.93 mJy). 

\begin{figure*}[htp]
\centering
\includegraphics[width=\textwidth]{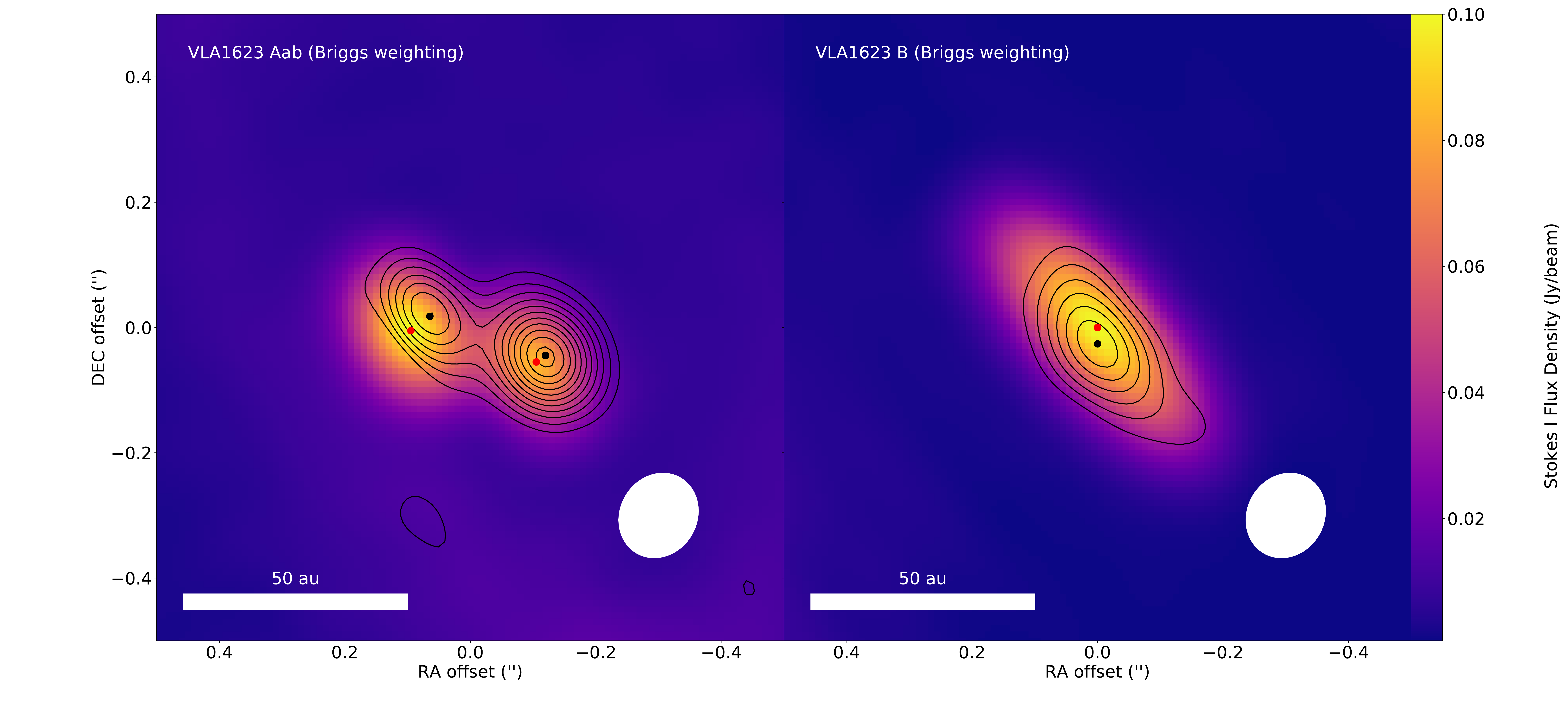}
\caption{(\textit{left}): Briggs weighted image of the 872 $\mu$m total continuum emission from VLA 1623Aab (colorscale) and of the polarized intensity (contours; contours start at 10$\sigma_{PI}$ and increment by 5$\sigma_{PI}$. The location of the peak of the total continuum emission is shown as a red dot; that of the polarized intensity is shown as a black dot. The scale-bar indicates 50 au. The synthesized beam is indicated in the lower right. \textit{(right)}: same as left, except for VLA 1623B.}
\label{fig:vla1623pi_i_comp}
\end{figure*}

One feature common to both the Briggs and the naturally weighted maps of the emission from VLA 1623A and VLA 1623B is that the polarization percentage is higher on one side of the source than on the other. Indeed, modeling the sources as elliptical Gaussians, the emission is more polarized on one side of line defining the major axis than the other. In VLA 1623Aa and VLA 1623Ab, the emission is more polarized to the north-west than to the south-east, and in VLA 1623B the emission is more polarized to the south-east than the north-west. This manifests as a $\sim$ 20-40 milliarcsecond offset of the peak of the polarized intensity away from the corresponding peak of the total intensity for each of VLA 1623Aa, VLA 1623Ab, and VLA 1623B. 
Figure \ref{fig:vla1623pi_i_comp} shows the relative astrometry of the peaks of the polarized emission and of the total emission. 
Given the high signal-to-noise
 ratio of the data with peak polarized intensity at $\gtrsim 40 \sigma_{PI}$ and peak total intensity at $\gtrsim 100 \sigma_{I}$, the offset of the peaks of the polarized intensity and total intensity is significant at the $\sim 8\sigma$ level for VLA 1623Ab and VLA 1623B and at the $\sim 11\sigma$ level for VLA 1623Aa. Neither VLA 1623W nor L1527 display such an offset, perhaps due to their extreme inclination.

To analyze and show the extended emission of VLA 1623CBdisk in more detail, we have produced an image of it alone without the A and B components (Figure \ref{fig:vla1623cb}). To produce this image, we performed image-plane fitting of the components of VLA 1623A from the Briggs-weighted map using the CASA task \texttt{imfit}; we then Fourier transformed the deconvolved best-fit model and subtracted it from the original visibility data. The difference was then Fourier transformed with natural weighting and deconvolved to produce the image in Figure \ref{fig:vla1623cb}. Because the fitting was performed on the Briggs-weighted map, some of the more extended emission associated with the stars was not captured; thus there is a residual $\sim$ 20 mJy/beam associated with the emission from VLA 1623A that was not subtracted out. However, the bulk of the emission ($\sim 95\%$) was removed in the fitting and subtraction, leaving evidence for a cavity in the center of the circumstellar material. Because \texttt{imfit} can only fit elliptical Gaussians to images, to estimate the flux density of the circumstellar material we smoothed the residual data with a Gaussian taper of half-width $100$k$\lambda$, imaged it, and used \texttt{imfit} to fit the disk. We find that the material has a flux density of $\sim$ 1 Jy and an extent of roughly $2\farcs2 \times 1\farcs2$. We note that this size may be a lower limit, as the estimated maximum angular scale recoverable by these observations is $\sim 2\farcs3$. Furthermore, it is larger than the extent recovered by \cite{sa18}, which may be due to either comparatively lower sensitivity at 1.3 mm or to size-dependent grain transport effects, as seen in later Class II sources \citep[e.g., AS 209;][]{pe12}

The naturally weighted map of the L1527 disk (Fig. \ref{fig:l1527}) shows an extended disk well detected in both the Stokes I emission as well as the polarized continuum. The integrated polarization is at approximately the 1.5\% level, but the polarization fraction clearly varies with both the distance from the midplane and distance from the star. Indeed, near the midplane, the polarization fraction reaches a peak value of $\sim$ 3 \% at the peak of the Stokes I emission and declines to $\lesssim 1\%$ towards the edges of the detectable Stokes I emission. The peak polarization percentage and morphology is consistent with that found in the 1.3 mm polarized continuum by \cite{se15} (note that their polarization vectors are rotated by 90 degrees to show an inferred magnetic field). However, we find that as the distance from the midplane of the disk increases, the polarization fraction increases from the $\lesssim 1-3\%$ to above 10\%, which was unseen by \citet{se15} due to their relatively poorer surface brightness sensitivity. 

One salient feature of each individual protostellar source in both the VLA 1623 and the L1527 systems is that the polarization is relatively well-ordered within the individual compact sources. To capture this and form an effective polarization angle for the emission associated with each compact source, we follow \cite{co18} and compute the polarized-intensity weighted average, i.e.
\begin{equation*}
<\theta> = \frac{\sum_{i} PI_i \theta_i}{\sum_{i}PI_i}\\
\end{equation*}
where $PI_i$ is value of the linearly polarized intensity in pixel $i$ and $\theta_i$ is the polarization position angle in the same pixel. 
We compute this average within a circle of radius $0\farcs1$ or about $\sim 14$ au for each individual protostellar source -- VLA 1623Aa, VLA 1623Ab, VLA 1623B, VLA 1623W, and L1527 -- centered on the maximum of the polarization intensity image. 
The mean angles for each component are shown in Table \ref{tab:angles}. Table \ref{tab:angles} also shows the minor axis position angle for each component ($\psi$) and the difference between the minor axis position angle and the mean polarization position angle ($|\theta - \psi|$). The quoted uncertainties of the polarization angles represent the standard deviation of sampled polarization angles within the circle. Were the disk and polarization position angles uncorrelated, we would expect to find no significant difference between our sample distribution of offsets and the parent distribution: a uniform distribution of values between 0 and 90 degrees. Table \ref{tab:angles} suggests however that these offsets are not uniformly distributed, but rather clustered around 0. Following \cite{co18}, we compared the distribution of angular offsets $|\theta-\psi|$ to that of a uniform distribution of angular offsets using both the Kolmogorov-Smirnov and Anderson-Darling tests. The likelihood that the sample of offsets as shown in Table \ref{tab:angles} is drawn from a uniform distribution is $P \lesssim 2 \times 10^{-4}$, which leads us to conclude that this sample is not drawn from such a distribution. In fact, given the proximity of all the angles to zero, it is likely that the true distribution of angular offsets is peaked at zero degrees, implying that the polarization angle is preferentially aligned with the minor axis of these disks.

\begin{deluxetable}{cccc}
\label{tab:angles}
\tablecaption{Orientations of disk polarization \& minor axis}
\tablenum{3}
\tablehead{\colhead{Star} & \colhead{$\theta =$ Polarization Angle} & \colhead{$\psi =$ Minor Axis PA} & \colhead{$|\theta - \psi|$} \\ 
\colhead{} & \colhead{(degrees)} & \colhead{(degrees)} & \colhead{(degrees )} } 
\startdata
VLA 1623Aa &   126.3 $\pm$ 3.0   &   138 $\pm$ 11   &  11.7 $\pm$ 11.4 \\
VLA 1623Ab &  130.2 $\pm$ 0.5    &   139 $\pm$ 6    &  8.8 $\pm$ 6 \\
VLA 1623B &  132.8 $\pm$ 1.0     &   131 $\pm$ 0.5  &  1.8 $\pm$ 1.1 \\
VLA 1623W &  100.3 $\pm$ 0.6      &  100 $\pm$ 0.5   &  0.3 $\pm$ 0.8 \\
L1527 &  98.3 $\pm$ 0.6      &   91.8 $\pm$ 0.6    &  6.5	$\pm$ 0.8 \\
\enddata
\tablecomments{All uncertainties are statistical. The systematic polarization angle uncertainty is not reflected in the quoted uncertainties.}
\end{deluxetable}

\section{Discussion}  \label{sec:disc}
Our high-resolution polarimetric observations of the young protostellar systems VLA 1623 and L1527 inform both how the polarization is produced in young Class 0 systems as well as on how these young systems form. We discuss each of the implications of our observations in turn.

\subsection{Polarization of the disk emission}
\label{ref:polsubsection}

The interpretation of our polarimetric observations is complicated by the fact that there are (at least) three different potential origins for the polarization--  direct emission from 
magnetically-aligned dust \citep{la07}, self-scattering of the thermal emission \citep{ka15,ya16a}, and direct emission from dust aligned by radiative flux \citep{ta17}-- that are at least partially degenerate with respect to the polarization morphology and spectrum they predict. 
In Section \ref{sec:results}, we presented our observations of the 872 $\mu$m thermal continuum emission from two young protostellar systems, VLA 1623 and L1527. In these two systems, there are six distinct structures: the individual sources that we refer to as VLA 1623Aa, VLA 1623Ab, and VLA 1623B, the extended disk of VLA 1623W, the extended circumbinary emission of VLA 1623CBdisk, and the disk surrounding L1527. We detect polarized emission from each object at a level of about 2-3\% of the local surface brightness and $\lesssim 1\%$ of the integrated flux density of each source. We discuss these different structures separately below.

We find compact polarization structures toward VLA 1623Aa, VLA 1623Ab, VLA 1623B, VLA 1623W, and L1527 that are consistent with the presence of disks. In each of VLA 1623Aa, VLA 1623Ab, VLA 1623B, and VLA 1623W, the polarization angle of the emission within $\sim$15 au of the peak is within a few degrees of the minor axis position angle; indeed, in these cases our measurements are consistent with the minor axis and polarization axis coinciding. In the disk associated with L1527, the minor axis  and polarization angles are slightly more misaligned, at $\sim 6$ degrees. The polarization angles of both VLA 1623B and (combined) VLA 1623Aab at 1.3 mm were found to be $\sim 130$ degrees \citep{sa18}, which is consistent with the results listed in Table \ref{tab:angles}. While \citeauthor{sa18} were able to resolve  VLA 1623B, they only marginally resolved VLA 1623Aab and could not separate it into two individual sources. In none of these sources do we observe any obvious curvature of the polarization structure or deviation 
away from the mean position angle. In fact, the geometry of the polarization may provide a powerful constraint for the origin of the emission in these systems.

Each of the different polarization mechanisms predicts polarization morphologies, depending principally on the optical depth $\tau$ of the emission and the disk inclination $i$. Were scattering to dominate the polarization signal of a disk with at least a moderate inclination angle, the polarization position angle would be parallel to the disk's minor axis essentially independent of $\tau$ interior to the outer edge of the disk. If one of the alignment mechanisms (i.e., magnetic alignment or alignment by radiative flux) is at play, optical depth significantly impacts the detected polarization morphology. The polarization angle can flip by $\sim$ 90 degrees due to differences in the disk layers (and, consequently, temperatures) effectively probed by the two orthogonal polarizations along the line of sight for an optically thick disk, as discussed below. 

As an example, suppose that a toroidal magnetic field aligns the grains in an inclined disk. In the optically thin limit, we would expect to see a radial polarization pattern with vectors pointing away from the center of the inclined disk. If, however, the disk is optically thick towards the center and optically thin at the outskirts, the polarization morphology is more complex and depends on both the vertical and radial temperature profiles in the disk. We would still see in the outer disk (i.e., in the optically thin regions) a radial pattern, but in the inner disk we would see deeper into the disk in the polarization parallel to the short axis of the grains than in the polarization parallel to the long axis due to the difference in extinction. Since the $\tau=1$ surface for the polarization parallel to the short axis is deeper into the disk, it could be hotter, especially for actively accreting disks, so the polarization could switch to a more azimuthal pattern, changing by $\sim 90$ degrees, absent of any change in the grain alignment. This polarization reversal due to differential extinction is a feature of both alignment mechanisms if there is a transition from low optical depth to high optical depth in a single source and if the temperature profile of the emitting grains declines with radius and height from the mid-plane. 

Given the information laid out above, we will now discern the most likely polarization mechanism in these protostellar sources.
Supposing that a toroidal field aligned grains in any of the relatively less-inclined structures in VLA 1623Aa, VLA 1623Ab, or VLA 1623B, the resulting emission in the optically thin limit would be purely radially polarized, and we would either observe such a polarization pattern or depolarization by beam smearing would suppress the polarization signal in the Q and U maps, particularly near the center of the disk. It is plausible that grains aligned with a poloidal magnetic field  would produce such a spatially uniform polarization. However, the pattern itself will depend on the degree to which the field perpendicular to the midplane ($B_z$) dominates over the radial field ($B_r$). If $B_z$ dominates where the polarization originates, then the grains would align their long axes parallel to the disk midplane. For significantly inclined disks as in VLA1623B, the grains would produce polarization parallel to the major axis if the disk is optically thin but perpendicular to the major axis (i.e., parallel to the minor axis) if the emission is optically thick due to the differential optical depth effect discussed above \citep[also as discussed in context of HH212; ][]{ya17,le18}. There is no indication that grains aligned by radiative flux are responsible for this emission. In this case, the grains would produce an azimuthally symmetric polarization pattern which would be beam-averaged out near the center \citep[as apparently happens in the case of Band 3 polarization of HL Tau; ][]{ka17} inconsistent with our strong detection near the center.  In addition to the morphology of the polarized emission, the relative offset of the peak of the polarized emission from the peak of the total emission that we see in each of VLA 1623Aa, VLA 1623Ab, and VLA 1623B is a clue. \cite{ya17} show that such an offset can be produced by self-scattering of the thermal dust emission. In a partially optically thick, inclined disk where the dust has not completely settled to the midplane, there is a difference in the effective inclination angle between the sides of the disk nearer and farther from the observer. 
Consequently, the emission from the side nearer to the observer is more polarized than that from the side further from the observer. Such an offset has previously been observed in the young early-type systems HH111 \citep{le18} and HH80-81 \citep{gi18} and inferred to be the signature of self-scattering in these systems somewhat more massive than VLA 1623. Neither of the alignment mechanisms clearly predict such an asymmetry as we observe. Given the clues from the polarization morphology as well as the offset of the polarized intensity from the total intensity in each disk, we conclude that scattering 
likely dominates the polarization in  VLA 1623Aa, VLA 1623Ab, and VLA 1623B, though optically thick emission from grains aligned with a poloidal field is another plausible origin.

The polarization morphology of VLA 1623CBdisk, in contrast to the other disks we observed, shows a mostly azimuthal morphology. The details of the polarization morphology mostly agree between our maps at 872 $\mu$m and those of \cite{sa18} at 1.3 mm, except in a small patch northwest of VLA 1623A where the polarization at 872 $\mu$m is rotated by $\sim$ 10 degrees compared to that at 1.3 mm, with the 872 $\mu$m emission being polarized more azimuthally. \citeauthor{sa18} interpret the 1.3 mm polarization in the ring as possibly  originating in dust grains aligned with a magnetic field with an hourglass morphology. Since the polarization vectors we recover from our data are mostly consistent with those from \citeauthor{sa18}, this is a plausible explanation for the recovered morphology. However, the rotation in position angle in the part of the disk where the data differ may be due to scattering becoming more important at the shorter wavelengths, leading to more azimuthally polarized emission in that part of the disk.

The edge-on disks L1527 and VLA 1623W are substantially more inclined than the other disks we observed, so we must we interpret the alignment of their polarization vectors with their minor axes differently. In these cases, neither scattering nor alignment with a toroidal magnetic field can immediately be rejected. On the other hand, the fact that the polarization is observed parallel to the minor axis well away from the midplane of the disk  is hard to explain for grains aligned by a poloidal magnetic field, because such sight lines are likely optically thin, which would lead to polarization parallel to the disk plane, which is not observed. Therefore, it seems unlikely that emission from grains aligned by a poloidal field is responsible for the emission in L1527 or VLA1623W.
Likewise, the emission does not come from grains aligned by radiative flux. The edge-on disks VLA 1623W and L1527 have  maximum polarization near the disk center. This is inconsistent with alignment by radiative flux, since the polarization morphology and percentage would not be so spatially uniform. Specifically, as the grains align in this scenario, they spin rapidly, and the polarization signal towards the disk center should be minimal because the plane-of-sky projection of the rapidly spinning grains would be a circle, which contradicts the fact that the polarization reaches a maximum near the disk center in VLA 1623W and L1527. Therefore, the origin of the polarized emission in these systems is presumably either scattering or emission by grains aligned with a toroidal field, although more detailed modeling that treats the optical depth effects quantitatively is needed to draw a firmer conclusion. 

Because the polarization geometry depends on optical depth in two of the three mechanisms, the brightness temperatures of the systems is an important cross-check on our conclusions. We measure the brightness temperature from our maps via the formula

\begin{eqnarray*}
T &=& \frac{\lambda^2 S_{\nu}}{2 k_B \Omega}\\
\end{eqnarray*} 
where $\lambda$ the observing wavelength, $k_B$ the Boltzmann constant, $\Omega$ the beam solid angle, and $S_{\nu}$ is the flux density. We simplify to find 
\begin{eqnarray*}
T  &=& \frac{1.36 \lambda^2 S_{\nu}}{\theta^2}\\
\end{eqnarray*}
where $T$ is the temperature in Kelvin, $\lambda$ the wavelength in mm, $\theta$ the synthesized beam half-width in arcseconds, and $S_{\nu}$ the recovered surface brightness in mJy/beam.
The peak brightness temperatures associated with VLA 1623Aa, Ab, and B are $\sim$ 60-70 K, while that associated with L1527 is $\sim$ 45 K. These temperatures are expected to be 
a substantial fraction of the kinetic temperature of the gas within a beam-width of the central protostar ($\sim 20-30$ au, $T\sim 50-100$K; e.g., \citealt{li17}), indicating an optical depth averaged over the beam of $\tau \gtrsim 1$ for each source assuming isothermal emission along the line of sight, although this estimate clearly depends on the temperature profile of the emitting material within the beam. 

For VLA 1623Aab, there are no published measurements of the kinetic temperature profile of the circumstellar material at the requisite spatial resolution to confirm our assertion that the 872 $\mu$m emission is optically thick. Accordingly, we must use estimates based on typical values for other protostars to assert that the material is likely optically thick based on the brightness temperature. On the other hand, the radial temperature profile of the disk of L1527 has been estimated from observations of optically thick $^{13}$CO($J=2-1$) and C$^{18}$O($J=2-1$) emission from the disk \citep{van18}. The reconstructed kinetic temperature profile of the disk material yields a temperature of $\sim$ 50 K or higher at 10 au, between 30-40 K at 25 au, and 25-35 K at 50 au. Given the brightness temperature, this implies $\tau \sim 1 - 2$ interior to about 30 au ($\sim 0\farcs2$), where it starts to decline to $\tau \sim 0.3$ at 40 au, near the edge of where polarized emission is detected within the midplane. Therefore, we expect that in L1527 we are detecting polarization across the $\tau = 1$ bridge between optically thick and optically thin emission, with which we can severely limit the likelihood that an alignment mechanism is at work. 

If $\tau$ peaks at $\gtrsim 1$ and either of the alignment mechanisms is at work, we might expect to see significant spatial variation in the polarization signal in our sources. This is because the orthogonal linear polarizations probe different regions of (i.e., temperatures in) the disk for a given line-of-sight. Due to the spatial variation in surface density, the degree to which the optical depths in the two polarizations differ will also spatially vary. On the other hand, scattering displays no such variation, and, indeed, scattering is more effective at producing $\sim 1$ \% levels of polarization than magnetically aligned grains as optical depth increases, and its relative importance increases as disks become more edge-on \citep{ya17}. Given that \cite{co18} find some statistical evidence that the polarization angle of compact sources in their sample is clustered around the minor axis position angle and that we find the same for our sources, we conclude that it is likely that the Band 7 continuum polarization is likely dominated by self-scattering in many Class 0 systems that may harbor disks.

\subsection{Structure of the VLA 1623 System}

As a hierarchical system with both individual circumstellar and circumbinary disks, the VLA 1623 system provides an interesting laboratory for studies of disk dynamics in complicated systems. Consistent with both \cite{mu13} and \cite{sa18}, we find that the extended disk associated with VLA 1623A has an inclination of $\sim$ 55 degrees at a position angle of about 25 degrees as determined from the ratio of the major and minor axes of the best-fit Gaussian. Furthermore, this disk surrounds two sources, VLA 1623Aa and VLA 1623Ab; each of these components has been resolved and is well fit by an elliptical Gaussian that suggests an inclination in the range of $45-55$ degrees with position angles that are consistent with each other. One plausible physical scenario that would create this is a binary system with individual circumstellar disks that are co-planar that is orbited by a circumbinary disk. The potential cavity of radius $\sim 0\farcs 4$ observed in Figure \ref{fig:vla1623cb} is marginally consistent with this. \cite{al94} showed that binaries can dynamically carve cavities within circumbinary disks that orbit them. For a binary of semimajor axis $a$, the inner cavity radius of a circumbinary disk should be $2-3 \times a$, with corrections that depend on the binary eccentricity $e$ and mutual inclination of the binary orbit and disk $i$. Such a scenario may also explain the brightness variation by a factor of  $\sim$ three from the north to the south side of the ring; binaries can create asymmetric pressure bumps within circumbinary disks, leading to enhanced emission at certain azimuths in the disk \citep{pi11,va13,bo17}

Due to the large spatial and kinematic offset between VLA 1623AB and VLA 1623W, it has been suggested that VLA 1623W was ejected during a close interaction with the remainder of the system during the system's formation \citep{mu13}. The later evolutionary class of VLA 1623W is consistent with such a scenario, since much of the nascent circumstellar material originally associated with it would be lost in the ejection \citep{re00}.
Our observations can be used to put constraints on the ejection scenario, by estimating the three-dimensional velocity of VLA 1623W as well and determining if such an ejection velocity is consistent with the resolved disk structure. By combining the data from ALMA 1.3 mm continuum  observations from \cite{mu13} with ours, we estimate that the proper motion of VLA 1623W is given by

\begin{eqnarray*}
\mu_{\alpha}\cos \delta &=& -14.9 \pm 7 \textrm{ mas yr}^{-1}\\
\mu_{\delta} &=& -23.4  \pm 7 \textrm{ mas yr}^{-1} \\
\end{eqnarray*}
This proper motion is formally consistent with both the results we obtain for the proper motion of VLA 1623B and that found by \cite{sa18}. Our results indicate that the differential proper motion of VLA 1623W relative to VLA 1623B is 
\begin{eqnarray*}
\Delta \mu_{\alpha}\cos \delta &=& -6.9 \pm 7.2  \textrm{ mas yr}^{-1} \\ 
\Delta \mu_{\delta}  &=& 3.6 \pm 7.2 \textrm{ mas yr}^{-1} \\ 
\end{eqnarray*}
Therefore, the available data is consistent with VLA 1623W comoving with VLA 1623B, albeit with relatively large uncertainties. However, the relatively large kinematic offset of VLA 1623W from VLA 1623AB ($\sim$ 4 km/s) as well as the resolved structure of the VLA 1623W disk may put severe constraints on the ejection scenario. 
Assuming that VLA 1623W shares its origin with VLA 1623AB as a member of a cluster of young objects, we can obtain a first order estimate of the ejection velocity by treating the system as a three-body system of point sources. The ejection velocity of a third component of a triple system depends on the geometry of the close encounter that produced it as well as the component masses and the angular momentum distribution within the system. However, the ejection velocity $v_{\mathrm{eject}}$ is typically of order $\sim 15 D^{-1/2}_{\mathrm{closest}}$ km/s, where $D_{\mathrm{closest}}$ is the distance in au of closest approach between the companion and the other stars \citep{ar97}. Because the viscous timescale within the disk becomes very long ($\sim 10^{5-6}$ years, \citealt[e.g., ][]{pr81})  as the edge of the disk is approached, the disk radius of VLA 1623W provides a useful constraint on how close the approach may have been without serious dynamical truncation \citep{al94}. We measure a radius of  $r \sim 50$ au, which implies that $D_{\mathrm{closest}}  \gtrsim 100$ au \citep[e.g., ][]{al94}, so $v_{\mathrm{eject}}$ is of order 1.5 km/s. Were the component ejected at this speed, it would take $\sim 6$ kyr to reach its current position, if the ejection velocity were solely in the plane of the sky. The line-of-sightkinematic offset itself between the two parts of the system is $3-4$ km/s. In three-body interactions wherein spin-orbit coupling is negligible, ejection preferentially occurs orthogonal to the total angular momentum of the system \citep{sa74,va05}. Due to the extent of the circumbinary disk of VLA 1623A, the bulk of the angular momentum in the system should reside in the disks associated with VLA 1623AB. Therefore, in the ejection scenario, VLA 1623W should be ejected roughly within the plane of the circumbinary disk of VLA 1623A. If the ejection indeed took place within the plane of the circumbinary disk, simultaneously satisfying the plane-of-sky displacement and the constraint on the ejection velocity is very difficult.  However, because the multiple-body interaction is in general complicated and depends greatly on initial conditions, it is plausible that the ejection speed would be much greater than that determined by the approximation above. 
Treating the system as a more realistic mixture of extended sources would most likely exacerbate the problem: ejection velocities of third bodies in simulations of very young multiple systems and of gravitationally bound clumps within protostellar disks have found ejection velocities of order $\lesssim 1$ km/s \citep{re10,ba12,vo16}.

\subsection{Formation of the VLA 1623AabB System}

With VLA 1623Aab a binary that lies within the gap of a non-axisymmetric circumbinary disk and VLA 1623B lying just at the edge of the disk, our 872 $\mu$m image  bears a cursory resemblence to the 1.3 mm image of the young triple source L1448 IRS 3B presented by \citet{to16b}. \citeauthor{to16b} conclude that L1448 IRS 3B formed via gravitational fragmentation due to its morphology, a Toomre $Q$ value of $Q \lesssim 1$, and simple analytic modeling of the disk. Likewise, we may also estimate the stability of the disk surrounding VLA 1623Aab. 

We can estimate the circumbinary disks' stability by estimating the value of the Toomre $Q$ paremeter, given by 

\begin{eqnarray*}
Q &=& 2\frac{M_{\mathrm{central}}}{M_{d}} \frac{c_s /\Omega}{r}
\end{eqnarray*}
where $M_d$ is the circumbinary disk mass, $M_{\mathrm{central}}$ the central object (in this case, the central binary mass), $c_s$ the disk sound speed, $\Omega$ the rotational angular velocity, and $r$ the distance form the central binary.
In the optically thin limit, the disk mass can be estimated from the 872 $\mu$m flux density via 
\begin{eqnarray*}
M_d &=& \frac{D^2 F_{\nu}}{\kappa_{\nu}B_{\nu}(T_d)}\\
\end{eqnarray*}
where $D$ is the source distance, $F_{\nu}$ the flux density, $\kappa_{\nu}$ the opacity, and $B_{\nu}(T_d)$ the Planck function evaluated at the dust temperature, $T_d$. We adopt a dust opacity law given by
\begin{eqnarray*}
\kappa_{\nu} &=& 10.0 \left(\frac{\nu}{\textrm{1 THz}}\right)^{\beta}\textrm{cm}^2\textrm{g}^{-1}\\
\end{eqnarray*}
with $\beta = 1$ \citep{be90}. Assuming a gas-to-dust mass ratio of 100:1, we find that the total opacity is 0.034 cm$^2$/g. We assume $T_d = 30$ K, which, combined with the brightness temperature of the circumbinary disk, is consistent with optically thin 872 $\mu$m emission from the reservoir. With these assumptions, $M_d \sim 0.03$ M\textsubscript{\(\odot\)}. For $T_d \sim 30$ K, $c_s \sim 330 $ m/s. Assuming a central mass of $\sim$ 0.15 M\textsubscript{\(\odot\)}, we find that the value of the $Q$ parameter is $\sim$ 1.8 at a distance of 85 au from the central binary, indicating that, on average, the disk is marginally stable to gravitational collapse. Of course, if the disk is partially optically thick, this is likely an overestimate. Due to the factor of $\sim$ 3 surface brightness variation across the circumbinary disk, it is plausible that some parts of the disk are more susceptible to collapse than others. 

\section{Conclusions} 
We have presented sensitive, high-resolution ALMA observations of the polarized 872 $\mu$m dust continuum from two protostellar, disk-bearing systems, the multiple system VLA1623 in Ophiuchus and the isolated protostar L1527 in Taurus. Our conclusions are as follows.

\begin{itemize}
\item We have found that the emission from VLA 1623A, previously determined to be the host of a Keplerian disk, comprises an extended disk with a partial gap and a source of compact emission that breaks up into two separate sources at high resolution. We interpret the separate sources as a proto-binary within the VLA 1623A system and note that VLA 1623 appears to be the first kinematically confirmed circumbinary disk for a Class 0 source.

\item We spatially resolved the disk of the distant companion VLA 1623W for the first time and find it to be a predominantly edge-on disk of radius $\sim$ 50 au.
\item The polarization signal of the circumbinary disk around VLA 1623Aab is strongly detected. Its polarization properties at 872 $\mu$m are mostly consistent with those found by \cite{sa18} at 1.3 mm.
\item The polarization signal of each of VLA 1623Aa, VLA 1623Ab, VLA 1623B, VLA 1623W, and L1527 is strongly detected, and their polarization angle is spatially constant. Given the systems' inclinations, the polarization angle distribution alone is consistent with an origin in optically thick emission from grains aligned with a poloidal field (VLA 1623Aa, VLA1623Ab, and VLA 1623B) optically thin emission from grains aligned with a toroidal field (VLA 1623W and L1527) or purely through scattering in either optical depth limit (all sources). In particular, we find evidence against polarization from grain alignment with respect to the radiative flux in all sources, particularly the two most edge-on disks, L1527 and VLA1623W.

\item In addition to the polarization angle distribution, our data provide other constraints on what mechanism prevails in determining the polarization. We have detected significant spatial offsets between the peak of polarized intensity and total intensity in the VLA 1623 system. In each of VLA 1623Aa, VLA 1623Ab, and VLA 1623B, the polarized intensity peak is displaced from the total intensity peak roughly along the minor axis. This is predicted to occur if the origin of the polarization is self-scattering in the optically thick limit in an inclined disk wherein dust has not yet settled to the midplane. No other mechanism naturally accounts for this offset. Since the brightness temperatures of the compact components of VLA 1623 are likely a significant fraction of the kinetic temperature, and since the brightness temperature of L1527 is close to the measured kinetic temperature of the gas in the midplane, it is likely that the emission from these two systems is optically thick. Given that and the sum of the morphological evidence, we find it very likely that scattering dominates the polarization signal in each source.

\item With our results combined with recent results on the Band 7 polarization of Class 0 disks, we find it likely that Band 7 polarization in a majority of these systems is caused by self-scattering and yields minimal information about the magnetic field. If this is the case, it may however be useful as a way to probe the size distribution of the dust population in these young systems.

\end{itemize}
\acknowledgments
{\centering ACKNOWLEDGMENTS \par}\
We thank the anonymous referee for their helpful comments. RJH thanks Nicholas Stone for useful conversations relating to three-body dynamics. ZYL is supported in part by NSF AST-1313083 and AST-1716259 and NASA NNX14AB38G. WK was supported by Basic Science Research Program through the National Research Foundation of Korea (NRF-2016R1C1B2013642). This research made use of APLpy, an open-source plotting package for Python \citep{ro12}
This paper makes use of the following ALMA data: ADS/JAO.ALMA\#2015.1.00084.S. ALMA is a 
partnership of ESO (representing its member states), NSF (USA) and NINS (Japan), together with 
NRC (Canada), NSC and ASIAA (Taiwan), and KASI (Republic of Korea), in cooperation with the 
Republic of Chile. The Joint ALMA Observatory is operated by ESO, AUI/NRAO and NAOJ. The NRAO is a facility of the National Science Foundation
operated under cooperative agreement by Associated Universities, Inc.
\software{CASA (v4.7.0; \citealt{mc07}), Matplotlib (http://dx.doi.org/10.1109/MCSE.2007.55)}

\end{document}